\documentclass[10pt]{article}
\usepackage{graphicx}

\begin{document}
\title{Electronic Enhancements in the Detection of
Gravitational Waves by Metallic Antennae}
\author{Y.N. Srivastava and A. Widom \\
Physics Department\\
Northeastern University, Boston MA 02115 USA \\
and \\
Physics Department \& INFN \\
University of Perugia, Perugia It \\
\\
G. Pizzella \\
Physics Department \& INFN\\
University of Rome II, Rome Italy}
\date{}
\maketitle
\begin{abstract}
For mechanical Weber gravitational wave antennae,
it is thought that gravity waves are weakly converted
into acoustic vibrations. Acoustic vibrations in metals
(such as Aluminum) are experimentally known to be
attenuated by the creation of electron-hole
pairs described via the electronic viscosity. These final
state electronic excitations give rise to gravitational wave
absorption cross sections which are considerably larger
(by four orders of magnitude) than those in previous theories
which have not explicitly considered electronic excitations.
\end{abstract}

\section{Introduction}

Weber gravitational wave antennae are designed to weakly convert
gravitational waves into acoustic
vibrations\cite{Weber1,Weber2}. The damping of acoustic vibrations
ultimately heat the detector accounting for the energy lost by the
incident gravitational disturbance. The total cross section for the absorption
of the gravitational wave is (of course) very small due to the weak
nature of the gravitational forces. If \begin{math} g \end{math}
denotes an incident graviton which is converted into a phonon
\begin{math} \phi \end{math},
then the first stage of graviton absorption is thought to
involve the conversion
\begin{equation}
g\to \phi .
\label{graviton_decay}
\end{equation}
In the second stage of absorption, the phonon decays.
In metallic samples (such as Aluminum) at not too high a temperature,
the dominant mode of decay of the phonon is into an
\cite{phonon_decay1,phonon_decay2}electron-hole pair
\begin{equation}
\phi \to e^- +e^+.
\label{phonon_decay}
\end{equation}
The decay in Eq.(\ref{phonon_decay}) is known to be true in virtue
of experimental work on sound wave attenuation in metals.
The final electronic excited states are described
in terms of an effective ``electronic viscosity''. Taking
Eqs.(\ref{graviton_decay}) and (\ref{phonon_decay}) together,
one finds that the gravitational wave decays into electronic
particle hole pairs via the process
\begin{equation}
g\to \phi  \to e^- +e^+
\label{g_process}
\end{equation}
wherein the acoustic phonon \begin{math} \phi  \end{math} enters
into the absorption process as a virtual (resonant) intermediate
state. Also possible is the {\em direct} conversion of the graviton
into an electron hole pair
\begin{equation}
g\to e^- +e^+ .
\label{ge_process}
\end{equation}

At first glance, it might appear that the electron-graviton
coupling would be much weaker than the phonon-graviton coupling
since the mass of the vibrating nuclei is much larger than the
mass of the electrons. However, as will be discussed in detail below,
the graviton couples into the {\em pressure} and {\em not} directly
into the energy density.
The spatial (pressure) part of the space-time stress tensor
\begin{math} T_{ij}=P_{ij} \end{math}
directly couples into a gravitational wave while the time
(energy density) part of the space-time stress tensor
\begin{math} T_{00}=\epsilon \end{math} couples into
the Newtonian part of gravity\cite{newt98}.

The electron contribution to an earthly (ordinary) condensed
matter pressure tensor \begin{math} P_{ij} \end{math} dominates
the nuclei contribution. The absorption rate of gravitational
disturbances are considerably enhanced by the large kinetic energy
and high density of final electronic states. A similar view has been
discussed in some previous work\cite {Weber3,Prep1}. Nevertheless, the
conventional wisdom is that the nature of the acoustic wave decay rate
has little to do with the total absorption rate of the gravitational wave,
provided that the former is much larger than the latter
\cite{Weinberg72,Ruffini74}. Our purpose is to exhibit the flaw in
the above conventional argument by providing a counter example.
The fact that the electrons dominate the coupling to both
the phonons and the gravitons considerably enhances the total cross
section in metallic antennae (such as Aluminum).

In Sec.2, the gravitational wave propagator will be discussed.
It will be shown that the ``self energy'' part of the propagator
(as a fourth rank tensor ``stress to strain ratio'') can be viewed
as a dynamical elastic tensor Young's modulus for distorting space-time.
In Sec.3, the gravitational wave will be treated as a weak
curvature strain on a flat space-time background. The dynamical
elastic moduli of metallic antennae will be explored.
In Sec.4, the amplitude for scattering a gravitational
wave off a metallic antenna will be computed in the mass quadrupole
approximation. For solid metals, the final total cross section
will be described in terms of the dynamical Lam\'e elastic coefficient
\begin{math}\mu \end{math}. In Sec.5, the microscopic quantum mechanical
expression for the gravitational wave total cross section will be
exhibited. The mathematical derivations rely on the Kubo formulae
(in condensed matter physics) for the transport
coefficients associated with phonon damping.
In Sec.6, the dispersion relations for the dynamical elastic
coefficients will be discussed. The results will be
applied to the subtracted dispersion relations for the dynamic Lam\'e
coefficient and the viscosity. The electronic viscosity enhancement
of the elastic Lam\'e coefficient will be explicitly exhibited.
In Sec.7, the Feynman diagrams for the absorption
of a graviton will be explored and the enhancement of the absorption
due to electronic coupling will be computed. In the concluding
Sec. 8, we present a brief summary of our results as well as
directions for future work.

\section{Gravitational Propagators}

In relativity one considers (in general) a background curved
space-time metric for the local proper time interval
\begin{equation}
-c^2d\tau ^2=g_{\mu \nu }dx^\mu dx^\nu
\label{GR1}
\end{equation}
which arises from a solution of the Einstein field equations
\begin{equation}
R_{\mu \nu }-\frac{1}{2}g_{\mu \nu }R
=\left(\frac{8\pi G}{c^4}\right)T_{\mu \nu}.
\label{GR2}
\end{equation}
Upon varying by \begin{math} \delta T_{\mu \nu}  \end{math}
the ``stress'' applied to space-time
\begin{equation}
T_{\mu \nu}\to T_{\mu \nu}+\delta T_{\mu \nu}
\label{GR3}
\end{equation}
one finds a response in the distortion (i.e. ``strain''
\begin{math} (1/2)\delta g_{\mu \nu} \end{math}) of space-time
by requiring a neighboring solution to Eq.(\ref{GR2}); i.e.
\begin{equation}
g_{\mu \nu}\to g_{\mu \nu}+\delta g_{\mu \nu}.
\label{GR4}
\end{equation}
The response in the strain due to the application of a stress
defines the gravitational wave propagator
\begin{math} D_{\mu \nu \lambda \sigma } \end{math} via
\begin{equation}
\delta g_{\mu \nu }(x)=\int D_{\mu \nu \lambda \sigma }(x,x^\prime )
\delta T^{\lambda \sigma }(x^\prime )d\Omega ^\prime
\label{GR5}
\end{equation}
wherein \begin{math} d\Omega =\sqrt{-g}d^4x \end{math}.

Let us decompose the stress into a source
\begin{math} \delta T_{ext}^{\lambda \sigma } \end{math}
{\em external} to the antenna and a stress {\em induced}
in the antenna
\begin{math} \delta T_{ind}^{\lambda \sigma } \end{math}
by the strain
\begin{math} (1/2)\delta g_{\mu \nu }  \end{math}. In total
\begin{equation}
\delta T^{\lambda \sigma }= \delta T_{ext}^{\lambda \sigma }
+ \delta T_{ind}^{\lambda \sigma } .
\label{GR6}
\end{equation}
The full propagator describes the response in the strain to
an ``external'' stress
\begin{equation}
\delta g_{\mu \nu }(x)=\int \Delta_{\mu \nu \lambda \sigma }(x,x^\prime )
\delta T_{ext}^{\lambda \sigma }(x^\prime )d\Omega ^\prime ,
\label{GR7}
\end{equation}
while the ``induced'' stress responds to the strain via the elastic
response function
\begin{equation}
\delta T_{ind}^{\lambda \sigma }(x^\prime )=\int
\Pi^{\lambda \sigma \alpha \beta }(x^\prime ,x^{\prime \prime })
\delta g_{\alpha \beta }(x^{\prime \prime })
d\Omega^{\prime \prime } .
\label{GR8}
\end{equation}
We note in passing that Eqs.(\ref{GR5})-(\ref{GR8}) imply Dyson's
Eq.(\ref{GR9}) below; i.e.
\begin{eqnarray}
\Delta_{\mu \nu \lambda \sigma }(x,x^{\prime \prime \prime })
&=&D_{\mu \nu \lambda \sigma }(x,x^{\prime \prime \prime })
\nonumber \\ &+& \int \int
D_{\mu \nu \alpha \beta }(x,x^{\prime })
\Pi^{\alpha \beta \xi \eta }(x^\prime ,x^{\prime \prime })
\Delta_{\xi \eta \lambda \sigma }
(x^{\prime \prime },x^{\prime \prime \prime })
d\Omega ^\prime d\Omega ^{\prime \prime }\ \ \
\label{GR9}
\end{eqnarray}
wherein the dynamic elastic coefficients
\begin{math} \Pi^{\alpha \beta \xi \eta } \end{math}
play the role of the gravitational wave propagator self energy.

Since the external stress radiates the incoming strain via
the relation
\begin{equation}
\delta g^{(in)}_{\mu \nu }(x)=\int D_{\mu \nu \lambda \sigma }
(x,x^{\prime })
\delta T_{ext}^{\lambda \sigma }(x^{\prime } )
d\Omega ^{\prime },
\label{GR10}
\end{equation}
it follows from Eqs.(\ref{GR7}), (\ref{GR9}) and (\ref{GR10}) that
the gravitational wave scattering equation takes the form
\begin{equation}
\delta g_{\mu \nu }(x)=\delta g^{(in)}_{\mu \nu }(x)+
\int \int D_{\mu \nu \alpha \beta }(x,x^\prime )
\Pi^{\alpha \beta \lambda \sigma }(x^\prime ,x^{\prime \prime })
\delta g_{\lambda \sigma }(x^{\prime \prime })
d\Omega^\prime d\Omega ^{\prime \prime }.
\label{GR11}
\end{equation}
In Eq.(\ref{GR11}), the total gravitational wave is the sum of an
incoming gravitational wave and an ``elastic'' scattered
gravitational wave
\begin{eqnarray}
\delta g_{\mu \nu }(x)&=&\delta g^{(in)}_{\mu \nu }(x)
+\delta g^{(el)}_{\mu \nu }(x)\nonumber \\
\delta g^{(el)}_{\mu \nu }(x)&=&
\int \int D_{\mu \nu \alpha \beta }(x,x^\prime )
\Pi^{\alpha \beta \lambda \sigma }(x^\prime ,x^{\prime \prime })
\delta g_{\lambda \sigma }(x^{\prime \prime })
d\Omega^\prime d\Omega ^{\prime \prime }.
\label{GR12}
\end{eqnarray}
From the imaginary part of the forward elastic scattering
amplitude one may compute the total cross section for the
wave to scatter off the antenna. Since the propagator
\begin{math} D_{\mu \nu \alpha \beta } \end{math}
contains the weak gravitational coupling strength
\begin{math} G  \end{math} to first order, it is
sufficient to employ the lowest Born amplitude for the
scattered gravitational wave
\begin{equation}
\delta g^{(el,Born)}_{\mu \nu }(x)=
\int \int D_{\mu \nu \alpha \beta }(x,x^\prime )
\Pi^{\alpha \beta \lambda \sigma }(x^\prime ,x^{\prime \prime })
\delta g^{(in)}_{\lambda \sigma }(x^{\prime \prime })
d\Omega^\prime d\Omega ^{\prime \prime }.
\label{GR13}
\end{equation}
To compute the scattered wave to lowest order
in \begin{math} G  \end{math} one needs to know
(i) the incoming gravitational wave
\begin{math} \delta g^{(in)}_{\lambda \sigma } \end{math},
(ii) the dynamic elastic coefficients
\begin{math} \Pi^{\alpha \beta \lambda \sigma }  \end{math}
of the antenna and (iii) the unperturbed (by the antenna)
gravitational wave propagator
\begin{math} D_{\mu \nu \alpha \beta } \end{math}.
Let us consider this in more detail.

\section{Gravitational Wave Antennae}

Consider a gravitational wave moving through a flat
space-time background in a reference frame in which an
antenna is at rest. In the transverse traceless gauge one may
describe the gravitational wave by a spatial {\em strain tensor}
\begin{math} u_{ij}({\bf r},t) \end{math} which enters into
the proper time as follows
\begin{eqnarray}
c^2d\tau ^2&=&c^2 dt^2-ds^2\nonumber \\
ds^2&=&|d{\bf r}|^2+2d{\bf r}\cdot {\bf u }\cdot d{\bf r}.
\label{GW1}
\end{eqnarray}
The definition of ``strain'' in terms of the spatial metric in
\begin{math} ds^2 \end{math} conforms to the standard usage in
elasticity theory\cite {Landau}. In terms of the traceless part of
the spatial pressure tensor
\begin{math} p_{ij}({\bf r},t) \end{math}
in the antenna,
\begin{equation}
{\bf p}={\bf P}-\frac{1}{3}{\bf 1}(tr{\bf P}),
\label{GW2}
\end{equation}
the gravitational wave equation (with a pressure source) reads
\begin{equation}
\left\{\frac{1}{c^2}\left(\frac{\partial }{\partial t}\right)^2
-\Delta \right\}{\bf u}({\bf r},t)
=\left(\frac{8\pi G}{c^4}\right){\bf p}({\bf r},t).
\label{GW3}
\end{equation}
The general solution of Eq.(\ref{GW3}) may be written
\begin{equation}
{\bf u}({\bf r},t)={\bf u}^{(in)}({\bf r},t)
+\left(\frac{2G}{c^4}\right)
\int \frac{{\bf p}({\bf r}^\prime ,t-R/c)}{R}d^3{\bf r}^\prime
\label{GW4}
\end{equation}
where \begin{math} R=|{\bf r}-{\bf r}^\prime |  \end{math}.
The incoming gravitational wave obeys
\begin{equation}
\left\{\frac{1}{c^2}\left(\frac{\partial }{\partial t}\right)^2
-\Delta \right\}{\bf u}^{(in)}({\bf r},t)=0.
\label{GW5}
\end{equation}

For a dynamical elastic antenna, the linear causal response
in the pressure \begin{math} {\bf p} \end{math} to a material
spatial strain \begin{math} {\bf u} \end{math} is described by
a non-local (in space and time) expression of the form
\begin{equation}
p_{ij}({\bf r},t)=-\int \int_0^\infty
{\cal G}_{ijkl}({\bf r},{\bf r}^\prime ,\bar{t})
u_{kl}({\bf r}^\prime ,t-\bar{t} )d\bar{t} d^3 {\bf r}^\prime .
\label{GW6}
\end{equation}
If the strain is turned on at a complex frequency
\begin{math} \zeta \end{math} with
\begin{math} {\Im m}\zeta >0\end{math} and if
\begin{equation}
{\bf u}({\bf r},t)=
{\Re e}\left\{e^{-i\zeta t} {\bf u}({\bf r};\zeta ) \right\},
\label{GW7}
\end{equation}
then the pressure response
\begin{equation}
{\bf p}({\bf r},t)=
{\Re e}\left\{e^{-i\zeta t} {\bf p}({\bf r};\zeta ) \right\}
\label{GW8}
\end{equation}
is described by a frequency dependent Young's modulus
\begin{math} {\cal Y} \end{math}. The explicit expression
for the pressure is
\begin{equation}
p_{ij}({\bf r};\zeta )=-\int
{\cal Y}_{ijkl}({\bf r},{\bf r}^\prime ;\zeta )
u_{kl}({\bf r}^\prime ;\zeta )d^3 {\bf r}^\prime
\label{GW9}
\end{equation}
where the dynamical Young's modulus obeys
\begin{equation}
{\cal Y}_{ijkl}({\bf r},{\bf r}^\prime ;\zeta )=\int_0^\infty
e^{i\zeta t}{\cal G}_{ijkl}({\bf r},{\bf r}^\prime ,t)dt.
\label{GW10}
\end{equation}
The scattering equation follows from  Eqs.(\ref{GW4})
and (\ref{GW7})-(\ref{GW9}); It is
\begin{eqnarray}
u_{ij}({\bf r};\zeta )&=&u^{(in)}_{ij}({\bf r};\zeta )+
u_{ij}^{(el)}({\bf r};\zeta )
\nonumber \\
u_{ij}^{(el)}({\bf r};\zeta )
&=& -\frac{2G}{c^4}\int \int
\left(\frac{e^{i\zeta R/c}}{R}\right)
{\cal Y}_{ijkl}({\bf r}^\prime ,{\bf r}^{\prime \prime };\zeta )
u_{kl}({\bf r}^{\prime \prime };\zeta )
d^3 {\bf r}^\prime d^3 {\bf r}^{\prime \prime }
\label{GW11}
\end{eqnarray}
where \begin{math} R=|{\bf r}-{\bf r}^\prime | \end{math}.

Note that Eq.(\ref{GW11}) is the flat space-time limit of
Eq.(\ref{GR12}) in thinly disguised form. Lowest order
perturbation in gravitational coupling constant
\begin{math} G \end{math} yields
\begin{equation}
u_{ij}^{(el,Born)}({\bf r};\zeta )=-\frac{2G}{c^4}
\int \int \left(\frac{e^{i\zeta R/c}}{R}\right)
{\cal Y}_{ijkl}({\bf r}^\prime ,{\bf r}^{\prime \prime };\zeta )
u^{(in)}_{kl}({\bf r}^{\prime \prime };\zeta )
d^3 {\bf r}^\prime d^3 {\bf r}^{\prime \prime }
\label{GW12}
\end{equation}
as the flat space-time limit of Eq.(\ref{GR13}).
Far away from the antenna, the scattered gravitational wave
for real frequency obeys
\begin{equation}
u_{ij}^{(el,Born)}({\bf r};\omega )\to
F_{ij}(\omega )\frac{e^{i\omega r/c}}{r}\ \ {\rm as}
\ \ r\to \infty
\label{GW13}
\end{equation}
where
\begin{equation}
F_{ij}(\omega )=-\left(\frac{2G}{c^4}\right)
\int \int e^{-i{\bf k}_f\cdot {\bf r}}
{\cal Y}_{ijkl}({\bf r},{\bf r}^\prime ;\omega+i0^+ )
u^{(in)}_{kl}({\bf r}^\prime ;\omega )
d^3 {\bf r} d^3 {\bf r}^\prime
\label{GW14}
\end{equation}
with the outgoing wave vector
\begin{math} {\bf k}_f=(\omega {\bf r}/cr)=
(\omega \hat{\bf r}/c) \end{math}.
If the incoming wave has the form
\begin{equation}
u^{(in)}_{ij}({\bf r};\omega )=
(\epsilon_{ij})_i e^{i{\bf k}_i\cdot {\bf r}},
\label{GW15}
\end{equation}
then the elastic scattering amplitude
\begin{equation}
(f|{\cal F}(\omega )|i)=(\epsilon_{ij})_f^*F_{ij}(\omega )
\label{GW16}
\end{equation}
follows from Eq.(\ref{GW14}) and (\ref{GW15}) to be
\begin{eqnarray}
(f|{\cal F}(\omega )|i)&=&-\left(\frac{2G}{c^4}\right)
\int \int d^3{\bf r} d^3{\bf r}^\prime \times  \nonumber \\
&\ &
\left\{
e^{-i{\bf k}_f\cdot {\bf r}}(\epsilon_{ij})_f^*
{\cal Y}_{ijkl}({\bf r},{\bf r}^\prime ;\omega+i0^+ )
(\epsilon_{kl})_i e^{i{\bf k}_i\cdot {\bf r}^\prime }\right\}.
\label{GW17}
\end{eqnarray}
The total cross section \begin{math} \sigma_i \end{math}
for a graviton (in state {\it i}) to be absorbed by the
antenna is determined by the imaginary part of the forward
scattering amplitude
\begin{equation}
\sigma_i(\omega )=\left(\frac{4\pi c}{\omega}\right)
{\Im m}(i|{\cal F}(\omega )|i).
\label{GW18}
\end{equation}
Eqs.(\ref{GW17}) and (\ref{GW18})  constitute the central result
of this section. The total cross section
\begin{math} \sigma \end{math} is determined by the dynamical
elastic Young's modulus \begin{math} {\cal Y} \end{math} of the
antenna.

\section{Mass Quadrupole Antennae}

If the wavelength of the incident graviton is large on the scale
of the size \begin{math} L \end{math} of the antenna,
\begin{equation}
(\omega /c)<<(1/L),
\label{QA1}
\end{equation}
then one may ignore the wave vectors
\begin{math} {\bf k}_{i,f}  \end{math}
in the integrals of Eq.(\ref{GW17}). This constitutes the mass
quadrupole approximation. For an antenna of volume
\begin{math} \Omega \end{math} and with
\begin{equation}
\Omega \chi_{ijkl}(\zeta ;\Omega )=
\int_\Omega \int_\Omega
{\cal Y}_{ijkl}({\bf r},{\bf r}^\prime ;\zeta )
 d^3{\bf r} d^3{\bf r}^\prime ,
\label{QA2}
\end{equation}
the mass quadrupole absorption cross section is
\begin{equation}
\sigma_i(\omega )\approx -\left(\frac{8\pi G\Omega }{c^3\omega }\right)
{\Im m}\left\{(\epsilon_{ij})_i^*
\chi_{ijkl}(\omega +i0^+;\Omega )(\epsilon_{kl})_i\right\}.
\label{QA3}
\end{equation}

Averaging over the initial gravitational wave polarizations
(i.e. helicity),
\begin{equation}
\sigma (\omega )=\frac{1}{2}\sum_{i=\pm}\sigma_i(\omega ),
\label{QA4}
\end{equation}
and defining the transverse dynamical elastic Lam\'e coefficient
as
\begin{equation}
\mu (\zeta ,{\bf n};\Omega )=
\frac{1}{4}\sum_{i^\prime =\pm }
\left\{(\epsilon_{ij})_{i^\prime }^*
\chi_{ijkl}(\zeta ;\Omega )(\epsilon_{kl})_{i^\prime }\right\}
\label{QA5}
\end{equation}
yields the total cross section
\begin{equation}
\sigma (\omega ,{\bf n};\Omega )
= -\left(\frac{16\pi G\Omega }{c^3\omega }\right)
{\Im m}\mu (\omega +i0^+ ,{\bf n};\Omega ),
\label{QA6}
\end{equation}
where \begin{math} {\bf n}  \end{math} denotes a unit vector along the
incident propagation direction of the gravitational wave. We note in
passing that the dynamical Lam\'e coefficient
\begin{math} \mu (\zeta ,{\bf n};\Omega ) \end{math}
for a finite volume system may have frequency
\begin{math} \zeta \end{math}-poles corresponding to both transverse
and longitudinal acoustic modes. The reason is that when a transverse sound wave
reflects off the boundary surface, a longitudinal sound wave may be produced with
finite amplitude and vice-versa.

We now presume that a large antenna can be represented as a
poly crystalline ``isotropic'' elastic system. The absorption
cross section shall nevertheless depend
on the incident propagation direction of the gravitational wave if the
shape of the antenna is {\em not} spherical.

The imaginary part of the Lam\'e coefficient may also be expressed
in terms of the real part of the viscosity coefficient according to
\begin{equation}
-{\Im m}\left\{\mu(\omega +i0^+ ,{\bf n};\Omega )\right\}=
\omega {\Re e}\left\{\eta (\omega +i0^+ ,{\bf n};\Omega )\right\}.
\label{QA7}
\end{equation}
Thus, with the mass of the antenna denoted by
\begin{math} M=\bar{\rho} \Omega  \end{math},
Eq.(\ref{QA6}) reads
\begin{equation}
\sigma (\omega ,{\bf n};\Omega )
= \left(\frac{16\pi GM }{c^2}\right)
\frac{{\Re e}\{\eta (\omega +i0^+ ,{\bf n};\Omega )\}}
{c\bar{\rho} }\ .
\label{QA8}
\end{equation}
Eq.(\ref{QA8}) is the central result of this section. The absorption
cross section for a graviton is directly related to the ``kinematic
viscosity'' \begin{math} (\eta /\bar{\rho}) \end{math} whose value determines
the life time of the converted transverse phonon.

\section{Microscopic Expressions}

If one writes the microscopic dynamical Young's modulus as
\begin{equation}
{\cal Y}_{ijkl}({\bf r}{\bf r}^\prime ;\zeta )=
{\cal Y}_{ijkl}({\bf r}{\bf r}^\prime ;0)
-i\zeta {\cal H}_{ijkl}({\bf r}{\bf r}^\prime ;\zeta ),
\label{ME1}
\end{equation}
then the Kubo formula for the non-local (in space) and frequency
dependent viscosity response function is given by
\begin{equation}
{\cal H}_{ijkl}({\bf r}{\bf r}^\prime ;\zeta )=
\frac{1}{\hbar }\int_0^\beta \left\{\int_0^\infty e^{i\zeta t}
\left<p_{kl}({\bf r}^\prime ,-i\lambda )p_{ij}({\bf r},t)\right>
dt\right\}d\lambda ,
\label{ME2}
\end{equation}
where
\begin{equation}
\beta =\left(\frac{\hbar }{k_B T}\right).
\label{ME3}
\end{equation}
The viscosity is then
\begin{equation}
\Omega \eta (\zeta ,{\bf n};\Omega )=
\frac{1}{4}\sum_{i^\prime =\pm }
\left\{
\int_\Omega \int_\Omega (\epsilon_{ij})_{i^\prime }^*
{\cal H}_{ijkl}({\bf r}{\bf r}^\prime ;\zeta )
(\epsilon_{kl})_{i^\prime }
d^3{\bf r}d^3{\bf r}^\prime \right\}
\label{ME4}
\end{equation}

From Eqs.(\ref{GW2}), (\ref{ME2}) and  (\ref{ME4}) it
follows that the pressure integral
\begin{equation}
\int {\bf P}d^3{\bf r}=
\frac{1}{2}\int {\bf r} {\bf r}divdiv{\bf P}d^3{\bf r}.
\label{ME5}
\end{equation}
is of central importance. Eq.(\ref{ME5}) follows from twice
integrating by parts. On the other hand, we have from
conservation of mass
\begin{equation}
\frac{\partial \rho}{\partial t}+div{\bf J}=0
\label{ME6}
\end{equation}
and from conservation of momentum
\begin{equation}
\frac{\partial {\bf J}}{\partial t}+div{\bf P}=0
\label{ME7}
\end{equation}
that
\begin{equation}
\frac{\partial^2 \rho}{\partial t^2}=divdiv{\bf P}.
\label{ME8}
\end{equation}
From Eqs.(\ref{ME5}) and (\ref{ME8}) it follows that
\begin{equation}
\int {\bf P}({\bf r},t)d^3{\bf r}=
\frac{1}{2}\left(\frac{d}{dt}\right)^2
\int {\bf r} {\bf r}\rho ({\bf r},t)d^3{\bf r}.
\label{ME9}
\end{equation}

For the problem at hand, the mass quadrupole of the antenna
may be defined as
\begin{equation}
{\bf D}(t)=\int_\Omega \left(3{\bf r}{\bf r}-r^2{\bf 1}\right)
\rho ({\bf r},t)d^3{\bf r}.
\label{ME10}
\end{equation}
From Eqs.(\ref{GW2}), (\ref{ME9}) and (\ref{ME10}) it follows that
\begin{equation}
\int_\Omega {\bf p}({\bf r},t)d^3{\bf r}=\frac{1}{6}\ddot{\bf D}(t),
\label{ME11}
\end{equation}
while Eqs.(\ref{ME2}) and (\ref{ME11}) imply
\begin{eqnarray}
h_{ijkl}(\zeta ,{\bf n};\Omega )&=&
\frac{1}{\Omega }\int_\Omega \int_\Omega
{\cal H}_{ijkl}({\bf r}{\bf r}^\prime ;\zeta )
d^3{\bf r}d^3{\bf r}^\prime
\nonumber \\
&=&\frac{1}{36\hbar \Omega }
\int_0^\beta \left\{\int_0^\infty e^{i\zeta t}
\left<\ddot{D}_{kl}(-i\lambda )\ddot{D}_{ij}(t)\right>
dt\right\}d\lambda .
\label{ME12}
\end{eqnarray}
Finally, from Eqs.(\ref{ME4}) and (\ref{ME12}) it follows that
\begin{equation}
\eta (\zeta ,{\bf n};\Omega )=
\frac{1}{4}\sum_{i^\prime =\pm }
(\epsilon_{ij})_{i^\prime }^*
h_{ijkl}(\zeta ,{\bf n};\Omega )
(\epsilon_{kl})_{i^\prime }.
\label{ME13}
\end{equation}
Eqs.(\ref{ME10}), (\ref{ME12}) and (\ref{ME13}) yield a rigorously
exact microscopic expression (to lowest order in {\it G}) for
the total gravitational wave cross section in Eq.(\ref{QA8}).

\section{Dispersion Relations and Sum Rules}

The viscosity coefficient obeys the analytic dispersion relation
\begin{equation}
\eta (\zeta ,{\bf n};\Omega)=-\left(\frac{2i\zeta }{\pi }\right)
\int_0^\infty
\frac{\Re e\{\eta (\omega +i0^+ ,{\bf n};\Omega )\}d\omega }
{\omega^2-\zeta^2}
\ \ {\rm for}\ \ {\Im m}\zeta >0.
\label{D1}
\end{equation}
With a finite viscosity \begin{math} \eta (0,{\bf n};\Omega) \end{math}
and shear modulus \begin{math} \mu (0,{\bf n};\Omega ) \end{math},
the full dynamical Lam\'e coefficient reads
\begin{equation}
\mu (\zeta ,{\bf n};\Omega)=\mu (0,{\bf n};\Omega)
-i\zeta \eta (\zeta ,{\bf n};\Omega).
\label{D2}
\end{equation}
In the limit \begin{math} |\zeta |\to \infty \end{math},
Eqs.(\ref{D1}) and (\ref{D2}) imply
\begin{equation}
\frac{2}{\pi }\int_0^\infty
{\Re e}\{\eta (\omega +i0^+ ,{\bf n};\Omega )\}d\omega =
\left(\mu (\infty ,{\bf n};\Omega)-\mu (0,{\bf n};\Omega)\right).
\label{D3}
\end{equation}

The microscopic evaluation of the sum rule follows
from Eqs.(\ref{ME12}), (\ref{ME13}) and (\ref{D3}) via an equal
time commutator
\begin{eqnarray}
\Delta \mu &=& \mu (\infty ,{\bf n};\Omega)-\mu (0,{\bf n};\Omega)
\nonumber \\
&=& \frac{1}{2\pi }\sum_{i^\prime =\pm }{\Re e}
\int_0^\infty (\epsilon_{ij})_{i^\prime }^* h_{ijkl}(\omega +i0^+
,{\bf n};\Omega)
(\epsilon_{kl})_{i^\prime }d\omega \nonumber \\
&=&\frac{1}{72\pi \Omega }{\Re e}\left\{\left(\frac{i}{\hbar }\right)
\sum_{i^\prime =\pm }(\epsilon_{ij})_{i^\prime }^*
\left<\left[\ddot{D}_{ij},\dot{D}_{kl}\right]\right>
(\epsilon_{kl})_{i^\prime } \right\}.
\label{D5}
\end{eqnarray}
From Eq.(\ref{ME10}) the mass quadrupole quantum operator may be
written as
\begin{equation}
D_{ij}=\sum_a M_a\left(3r_{ai}r_{aj}-\delta_{ij}r_a^2\right).
\label{D6}
\end{equation}
wherein \begin{math} M_a \end{math} is the mass and
\begin{math} {\bf r}_a \end{math} is the position of the
\begin{math} a^{th} \end{math} particle in the gravitational
antenna. These particles include both the electrons and the nuclei
in the condensed matter. If we employ the Hamiltonian
\begin{math} H=K+U \end{math} which is the
sum of the kinetic energy \begin{math} K \end{math} and
the potential energy \begin{math} U \end{math}, i.e.
\begin{eqnarray}
H&=&\sum_a\left(\frac{p_a^2}{2M_a}\right)+
U(\ldots,{\bf r}_a,\ldots) \nonumber \\
&=&\sum_a\left(\frac{p_a^2}{2M_a}\right)+
\sum_{a<b} u_{ab}(r_{ab}),
\label{D7}
\end{eqnarray}
then the time rate of change of the quadrupole operator in
Eq.(\ref{D6}) reads
\begin{equation}
\dot{D}_{ij}=\sum_a
\left\{3(r_{ai}p_{aj}+p_{ai}r_{aj})-\delta_{ij} ({\bf r}_a \cdot
{\bf p}_a+{\bf p}_a \cdot {\bf r}_a)\right\}. \label{D8}
\end{equation}
Similarly,
\begin{eqnarray}
\ddot{D}_{ij}&=&\sum_a
\left\{\left(\frac{6p_{ai}p_{aj}}{M_a}\right)+
3(r_{ai}f_{aj}+r_{aj}f_{ai})\right\} \nonumber \\
&\ &\ \ -2\delta_{ij}\sum_a\left\{ \left(\frac{p_a^2}{M_a}\right)
+({\bf r}_a\cdot {\bf f}_a)\right\}, \label{D9}
\end{eqnarray}
where for the case of two particle potentials
\begin{equation}
\dot{\bf p}_a={\bf f}_a=-{\bf \nabla}_a U =-{\bf \nabla}_a
\sum_{b\ne a}u_{ab}(r_{ab}).
\label{D10}
\end{equation}
The equal time commutation relations are computed employing
\begin{equation}
\left[p_{ai},r_{bj}\right]=-i\hbar \delta_{ab} \delta_{ij}.
\label{D11}
\end{equation}
A very tedious computation employing Eqs.(\ref{D5})-(\ref{D11})
yields the sum rule
\begin{eqnarray}
\Delta \bar{\mu } &=& \frac{2}{\pi}\int_0^\infty
{\Re e}\bar{\eta }(\omega +i0^+;\Omega ) d\omega
\nonumber \\
&=& \frac{1}{15\Omega }
\left\{
10\left<K\right>+
4\left<\sum_{a<b}r_{ab}u_{ab}^\prime (r_{ab})\right>
+\left<\sum_{a<b}r_{ab}^2u_{ab}^{\prime \prime }(r_{ab})\right>
\right\},
\label{D12}
\end{eqnarray}
where \begin{math} \bar{\eta }(\zeta ;\Omega ) \end{math} represents
a spherical average over \begin{math} {\bf n}  \end{math} in
\begin{math} \eta (\zeta ,{\bf n};\Omega ) \end{math}.
For the case of a model Hamiltonian for which the potentials
are completely described by Coulomb's law,
\begin{equation}
U=e^2\sum_{a<b}\frac{z_az_b}{r_{ab}},
\label{D13}
\end{equation}
Eqs.(\ref{D12}) and (\ref{D13}) imply
\begin{equation}
\Delta \bar{\mu }=\frac{2}{\pi}
\int_0^\infty {\Re e}\bar{\eta }(\omega +i0^+;\Omega ) d\omega
=\left(\frac{10\left<K\right>-2\left<U\right>}{15\Omega }\right).
\label{D14}
\end{equation}
From the virial theorem for a collection of particles
(electrons and nuclei) interacting with pure Coulomb potentials, the
thermodynamic pressure obeys
\begin{equation}
P=\left(\frac{2\left<K\right>+\left<U\right>}{3\Omega }\right)
=\left(\frac{\Omega \epsilon +\left<K\right>}{3\Omega }\right),
\label{D15}
\end{equation}
while the energy per unit volume
\begin{equation}
\epsilon=\frac{\left<K\right>+\left<U\right>}{\Omega }.
\label{D16}
\end{equation}
Eqs.(\ref{D14})-(\ref{D16}) imply
\begin{equation}
\Delta \bar{\mu}=\frac{2}{\pi}
\int_0^\infty {\Re e}\bar{\eta }(\omega +i0^+;\Omega ) d\omega
=\left(\frac{36P-14\epsilon }{15 }\right).
\label{D17}
\end{equation}
Eqs.(\ref{D15})-(\ref{D17}) imply that
\begin{equation}
\Delta \bar{\mu }=\left(\frac{14}{15}\right)\bar{t}
-\left(\frac{2}{5}\right)P,
\label{D18}
\end{equation}
where the kinetic energy per unit volume
\begin{equation}
\bar{t}=\left<K\right>/\Omega .
\label{D19}
\end{equation}
For pressures of the order of one atmosphere (or less)
one finds that \begin{math} \bar{t}>>P  \end{math}. Thus
\begin{equation}
\frac{\Delta \bar{\mu}}{\bar{\mu}}
\approx \left(\frac{14\bar{t}}{15\rho v_s^2}\right)
\label{D19}
\end{equation}
wherein \begin{math} \rho \end{math} and \begin{math} v_s  \end{math}
represent, respectively, the mass density and the shear sound wave
velocity. Note that the dynamic Lam\'e coefficient
\begin{math} \mu (\zeta ,{\bf n};\Omega ) \end{math} contains
\begin{math} \zeta  \end{math}-poles at modes determined by both
the longitudinal and transverse acoustic phonons.
Nevertheless, the zero frequency {\em static} Lam\'e coefficient
\begin{math} \bar{\mu }=\rho v_s^2 \end{math} depends only on the
shear (transverse) wave sound velocity \begin{math} v_s \end{math}.

The mean kinetic energy per unit volume is dominated by
the electrons. If \begin{math} n \end{math} represents the number
of electrons per unit volume, then a Thomas-Fermi uniform electron
distribution estimate of the kinetic energy per unit volume of the
electrons reads
\begin{equation}
\bar{t}\approx \left(\frac{3(3\pi ^2)^{2/3}}{10}\right)
\left(\frac{\hbar^2 n^{5/3}}{m}\right).
\label{D20}
\end{equation}
For example, an Aluminum atom has \begin{math} Z=13 \end{math}
electrons and an atomic weight of \begin{math} A=27 \end{math}
so that the number of electrons per unit volume
is given by \begin{math} n=(Z\rho /A\hat{M}) \end{math}.
Here, \begin{math} \hat{M} \end{math} is the atomic mass unit.
Thus, we find the estimate
\begin{equation}
\left(\frac{\Delta \bar{\mu }}{\bar {\mu }}\right)\approx 84
\ \ ({\rm Aluminum}).
\label{D21}
\end{equation}
The electronic contribution \begin{math} \Delta \bar{\mu } \end{math}
to the dynamic Lam\'e coefficient leads to an electronic
renormalization of the conversion of a graviton into a phonon as
will be discussed directly below.

\section{Enhancement Factor for Graviton Absorption}

\begin{figure}[tp]
\centering
\includegraphics[width=3in]{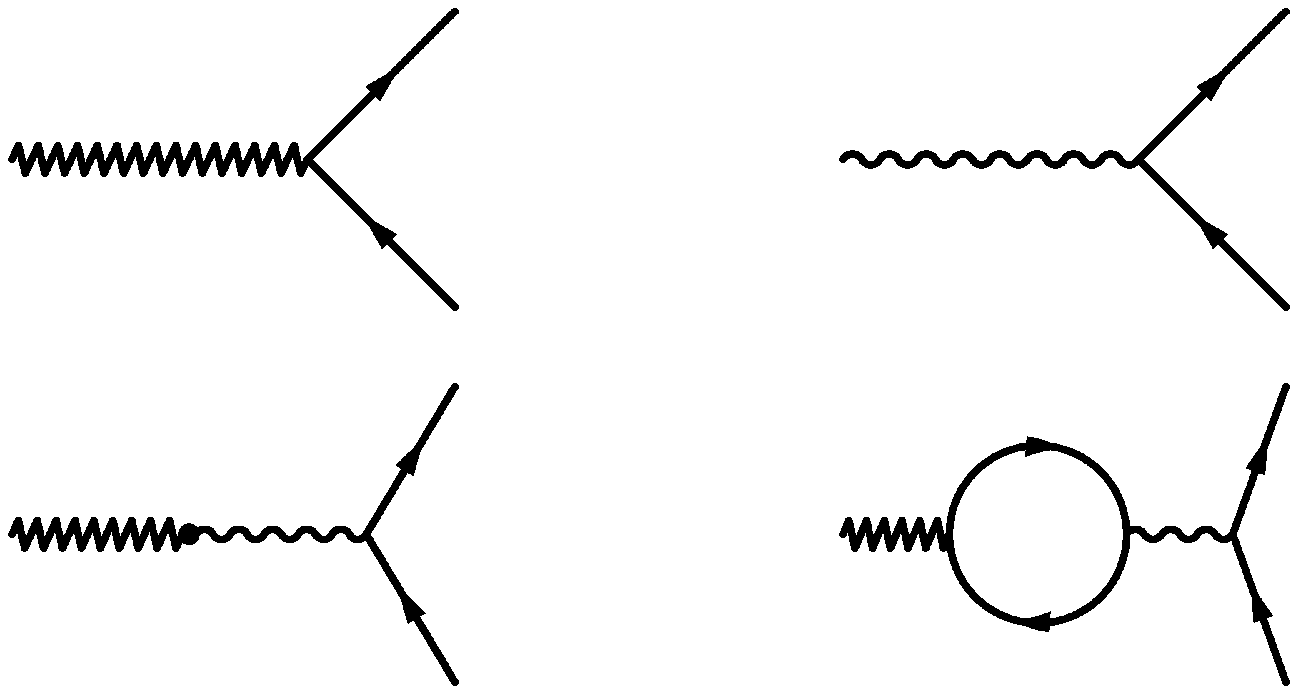}
\caption{Shown above are the following processes: (i) A graviton
decays into a particle hole pair $g\to e^-+e^+$. (ii) A phonon decays
into a particle hole pair $\phi \to e^-+e^+$. (iii) A graviton
again decays into a particle hole pair, but there is a conversion
to an intermediate resonant phonon $g\to \phi \to e^-+e^+$.
The conversion coupling strength is $\bar{\mu}$.
(iv) A graviton decays into a particle hole pair, again
with a conversion to an intermediate resonant phonon
$g\to [e^-e^+] \to \phi \to e^-+e^+$ employing an electron
loop coupling. The enhanced loop conversion coupling strength is
$\Delta \bar{\mu}$.}
\label{feynman}
\end{figure}

The Feynman diagrams for the physical processes here
considered are shown in Fig.\ref{feynman}. The absorption rates
are controlled by the material viscosity
\begin{math} \eta \end{math}
which for a metal (at reasonably low temperatures) is
determined by the motions of the electrons.
(i) If there is one graviton inside of a very large material
object of volume \begin{math} \Omega \end{math},
then the transition rate per unit time for a low frequency graviton
to be absorbed is equal to the flux per unit time per unit area
times the cross section; i.e.
\begin{equation}
\Gamma_g=\lim_{\omega \to 0}
\left(\frac{c\bar{\sigma} (\omega )}{\Omega }\right)
\label{en1}
\end{equation}
which, in virtue of Eq.(\ref{QA8}), reads
\begin{equation}
\Gamma_g=
\left(\frac{16\pi G }{c^2}\right)\eta
\ \ \ \ {\rm for}\ \ \ \ g\to e^-+e^+.
\label{en2}
\end{equation}
(ii) Similarly, the absorption of a phonon is also
controlled by the material transverse and bulk
viscosities, respectively, \begin{math} \eta \end{math} and
\begin{math} \eta_b \end{math}. For a phonon with wave
vector \begin{math} {\bf k}  \end{math} the transition per unit
time for absorption by the material is given by
\begin{eqnarray}
\Gamma_\phi (longitudinal) &=&
\left(\frac{|{\bf k}|^2}{\rho }\right)
\left(\eta_b +\frac{4\eta}{3}\right)
\ \ \ {\rm for}\ \ \ \phi \to e^-+e^+, \nonumber \\
\Gamma_\phi(transverse) &=&
\left(\frac{|{\bf k}|^2}{\rho }\right)\eta
\ \ \  {\rm for}\ \ \ \phi \to e^-+e^+ .
\label{en3}
\end{eqnarray}
(iii) In gravitational wave antennae the absorption of
a graviton takes place through the intermediate conversion
of a resonant phonon via
\begin{equation}
g\to \phi \to e^-+e^+.
\label{en4}
\end{equation}
The matrix element for the conversion
\begin{math} g\to \phi  \end{math}
is described by the Hamiltonian
\begin{equation}
H^\prime (g,\phi )=2\bar{\mu} \int ({\bf s}:{\bf u})d^3 {\bf r},
\label{en5}
\end{equation}
where (in the antenna) \begin{math} {\bf s}  \end{math} is the
strain due to the phonon and \begin{math} {\bf u} \end{math}
is the strain due to the graviton.
Eq.(\ref{en5}) describes the graviton conversions into both
longitudinal and transverse phonons. Nevertheless, only the
Lam\'e coefficient \begin{math} \mu  \end{math} enters into
the coupling. The reason is that the graviton is purely
transverse and traceless no matter which kind of phonon
it produces. (iv) An electronic enhancement of the graviton to
resonant phonon conversion process arises from the electron
loop process shown in Fig.(\ref{feynman}); i.e. the one loop
conversion process
\begin{equation}
g\to [e^-e^+]\to \phi
\label{en6}
\end{equation}
determines the coupling strength renormalization
\begin{equation}
\bar{\mu }\to \bar{\mu }+\Delta \bar{\mu },
\label{en7}
\end{equation}
wherein \begin{math} \Delta \bar{\mu }  \end{math} has been
evaluated in Eq.(\ref{D19}).

Since the absorption rate for gravitons is computed from the
absolute value squared of the matrix element, the central
point of this work is finally evident. The enhancement
of the graviton to resonant phonon detection process via
electronic loop viscosity is given by
\begin{equation}
{\cal E}=
\left|\frac{\bar{\mu}+\Delta \bar{\mu}}{\bar{\mu}}\right|^2 .
\label{en8}
\end{equation}
For the case of an Aluminum gravitational wave antenna,
Eqs.(\ref{D21}) and (\ref{en8}) imply a {\em large}
electronic induced detection enhancement; It is
\begin{equation}
{\cal E}\approx 7.2\times 10^3
\ \ ({\rm Aluminum}).
\label{en9}
\end{equation}
Such an enhancement allows for the experimental feasibility of
detecting gravitational radiation from sources within our own
galaxy.

\section {\textbf{Conclusions}}

In this work, we have considered in detail the dynamical response
produced in a metallic Weber bar by a gravitational wave. Our
treatment differs from all previous analyses in several major
aspects. First, departing from tradition, we follow the Einstein
equations so that a (transverse, traceless) gravitational wave is
directly coupled to the (transverse, traceless) part of the
pressure. Secondly, observing that in a metal (at not too high a
temperature) electrons (not the phonons) provide the bulk of the
pressure. Previous estimates of the couplings based on direct
graviton to resonant phonon conversions need a major revision. In
fact, we find that the electron-hole pair loop contributions
significantly renormalize the relevant couplings. We have computed
this using a Thomas-Fermi model with uniform electron density
to estimate the average kinetic energy per unit volume of the electrons.
The enhancement factor is found in this way to be
\begin{math}{\cal E}\sim 7\times 10^3 \end{math}.
The true factor is likely to be somewhat higher since the assumption
of uniform electron density underestimates the kinetic energy of the
localized core electrons.

Our results can be employed to explain the solution to a
theoretical problem raised by experimental observations with the
room temperature resonant detectors of the Maryland and Rome groups during the
SN1987A\cite{Amaldi,Aglietta1,Aglietta2}.
At this time the Rome and Maryland detectors showed
signals with energy of about 100 K correlated with the observation of
the neutrino detectors Mont Blanc, Kamiokande, Baksan and IMB. Using the
previous cross-section, a \begin{math} 100 K \end{math} signal requires
the total conversion into gravitational waves of about one thousand solar masses.
This would be impossible for an original star mass of
\begin{math} M\sim 20 M_{sun} \end{math}.
However, with the renormalized detection cross section of this work,
each \begin{math} 100 K \end{math} signal requires just one tenth of a
solar mass (still high but energetically possible) yielding important
information on collapse mechanisms. Also the rennormalized cross-section
yields a sensible astrophysical interpretation of the recent
result concerning experimental coincidences\cite{Astone} between the cryogenic
resonant detectors Explorer (CERN) and Nautilus (Frascati).
\bigskip
\par \noindent
We thankfully acknowledge useful comments, discussions and correspondences
on the subject matter of this paper with Professors N. Cabibbo, E. Coccia,
A. Degasperis, V. Ferrari, G. Immirzi, F. Palumbo and S. Weinberg.

\end{document}